\documentclass[12pt]{article}

\begin{document}

\begin{center}

{\Large {Sakharov's induced gravity on the AdS background: \\ SM scale as
inverse mass parameter of the Schwinger-DeWitt expansion}} \footnote{This is a slightly improved version of an article published on 09.09.2015 in Phys. Rev. D {\bf 92}, 065007.}

\vspace{1,5cm}

{Boris L. Altshuler}\footnote{E-mail addresses: baltshuler@yandex.ru $\,\,\,  \& \,\,\,$  altshul@lpi.ru}

\vspace{1cm}

{\it Theoretical Physics Department, P.N. Lebedev Physical
Institute, \\  53 Leninsky Prospect, Moscow, 119991, Russia}

\vspace{1,5cm}

\end{center}

{\bf Abstract:} One loop quantum effective action $W$ of scalar field 'living' on the AdS background of the Randall-Sundrum model is defined here by now popular way which excludes bulk UV divergencies; thus induced Planck mass is given not by the UV regularization parameter, like in Sakharov's pioneer work, but by the location of the UV-cut of AdS space. Resummation of Schwinger-DeWitt expansion of the action $W$ is performed by the novel 'auxiliary mass' method. The inverse mass squared parameter of this expansion is determined by the location of the 'visible' IR-brane of the RS-model. Obtained expression for induced vacuum energy density coincides with the independently calculated VEV of the stress-energy tensor. Corresponding potential in 4 dimensions possesses a non-trivial extremum which hopefully will permit us to stabilize IR brane and hence to fix the observed small value of mass hierarchy in analogy with the Coleman-Weinberg mechanism. It is demonstrated that naive equating of values of the induced Planck mass and vacuum energy density to those of the RS-model determines otherwise arbitrary constants of the model. A principle of quantum self-consistency generalizing this approach is proposed.

\vspace{0,5cm}

PACS numbers: 11.10.Kk, 04.50.-h

\newpage

\tableofcontents

\newpage

\section{Introduction}

\quad In 1967 Andrei Sakharov proposed to cut out the term of 'bare' Einstein Action in the gravity-matter Lagrangian

\begin{equation}
\label{1}
S=M_{Pl}^{2} \int R \sqrt{g}\, d^{4}x + S_{\it{matter}}
\end{equation}
and to induce this term from the one loop quantum fluctuations of scalar or other matter fields moving in external gravitational field \cite{Sakharov}. This approach, which Sakharov called "the theory of zero Lagrangian", was very promising and justifiably entered the textbooks (see e.g. \cite{Misner}). Surely now in string theory it looks somewhat trivial. Sakharov himself wrote in 1980th: "String theory is, at a new level, the realization of my old ideas concerning induced gravitation! I cannot refrain from feeling proud on this point" \cite{Gorky} (see in more detail in review \cite{Alt1991}). 

In Sakharov's approach induced gravitational constant is determined by the UV regularization parameter. In string theory it is given by string tension. In the present paper it will be expressed through the location of UV-cut of AdS space. Thus let us try to apply 'zero Lagrangian' approach in $(d+1)$-dimensional Randall-Sundrum model \cite{Randall} described by the metric

\begin{equation}
\label{2}
ds^{2}= \frac{1}{(kz)^{2}}[dz^{2}+{\tilde g}_{\mu\nu}(x)dx^{\mu}dx^{\nu}],
\end{equation}
$\mu,\nu = 0,1...d-1$, in what follows we use Euclidean signature, i.e. ${\tilde g}_{\mu\nu} = \delta_{\mu\nu}$ for the background AdS space. Also only zero modes of gravity field in $(d+1)$ dimensions will be considered, i.e. ${\tilde g}_{\mu\nu}$ is taken non-dependent on $z$. As ordinary integrals over $z$ are taken in the slice of AdS:

\begin{equation}
\label{3}
\epsilon \equiv z_{\it{UV}} < z < z_{\it{IR}} \equiv L,
\end{equation}
($L \simeq M_{\it{SM}}^{-1}$, where $M_{\it{SM}}$ is the Standard Model mass scale; $L \gg \epsilon, k^{-1}$; it is always possible to put $\epsilon = k^{-1}$ by the constant scale transformation of coordinates not changing metric (\ref{2})). However asymptotic boundary conditions of Green functions will be imposed not at $z=\epsilon$, but at the AdS horizon $z \to 0 $ (cf. e.g. \cite{Witten}, \cite{Mitra}). Bulk scalar curvature of space (\ref{2}) is

\begin{equation}
\label{4}
R_{\it{bulk}}^{(d+1)} = -d(d+1)k^{2} + k^{2}z^{2}{\tilde R}^{(d)}
\end{equation}
where ${\tilde R}^{(d)}$ is curvature in $d$ dimensions built with metric ${\tilde g}_{\mu\nu}(x)$.

In standard Kaluza-Klein approach Einstein Action in 4 dimensions (here - in $d$ dimensions) is obtained when extra coordinates are integrated out in Einstein term of the higher-dimensional action:

\begin{eqnarray}
\label{5}
S_{\it{bulk}}^{(d+1)} = M_{(d+1)}^{d-1}\int\, d{\it {vol}}_{(d+1)}\left[R^{(d+1)} - \Lambda \right] = \qquad \qquad \nonumber \\
\nonumber \\
= M_{(d+1)}^{d-1} \int \sqrt{{\tilde g}}\, d^{d}x\int_{\epsilon}^{L} \frac{dz}{(kz)^{(d+1)}} \left[ -d(d+1)k^{2} + k^{2}z^{2} {\tilde R}^{(d)} - \Lambda \right],
\end{eqnarray}
where $M_{(d+1)}$ is 'Planck mass' in $(d+1)$ dimensions, ${\tilde g}$ is determinant of metric ${\tilde g}_{\mu\nu}$, $\Lambda = - d(d-1)k^{2}$. Then negative vacuum energy in RS model in $(d+1)$ dimensions ($V^{(d+1)}_{\it{vac \, (RS)}}$) which is necessary to receive AdS space as a solution of Einstein equations, and Planck mass in $d$ dimensions ($M_{(d)\,{\it (RS)}}$) are given by well known expressions

\begin{equation}
\label{6}
V^{(d+1)}_{\it{vac\,(RS)}} = - d(d-1)k^{2}M_{(d+1)}^{d-1}, \qquad M^{d-2}_{(d){\it (RS)}} = \frac{M_{(d+1)}^{d-1}}{(d-2)k^{d-1}\epsilon^{d-2}}
\end{equation}
(in expression for Planck mass small correction due to upper limit $z=L$ in integral over $z$ in (\ref{5}) is omitted).

In this paper the attempt is made to receive constants (\ref{6}) from $S_{\it{matter}}$ as a quantum induced phenomenon. We consider one loop effective Action of scalar field $\Phi (z,x)$ 'living' in space (\ref{2}) and described by the minimal Action:

\begin{eqnarray}
\label{7}
S_{\it{matter}} = -\frac{1}{2} \int d{\it{vol}}_{(d+1)} \left[(\nabla\Phi)^{2}+m^{2}\Phi^{2} \right] = \qquad  \qquad  \nonumber \\
\nonumber \\
= -\frac{1}{2} \int \sqrt{{\tilde g}}\, d^{d}x\int_{\epsilon}^{L} \frac{dz}{(kz)^{(d+1)}} \left[ k^{2}z^{2} \Phi_{,z}^{2} + k^{2} z^{2} {\tilde{g}}^{\mu\nu}\Phi_{,\mu}\Phi_{,\nu} + m^{2} \Phi^{2} \right],
\end{eqnarray}
where the comma means an ordinary derivative, and $m$ is the mass of scalar field.

The conventional Schwinger-DeWitt (S-DW) or heat kernel expansion of the one loop effective action in higher powers of curvature and its derivatives on the flat background \cite{DeWitt}, \cite{Birrell}, \cite{BarvVilk}, \cite{Vass} can not be applied here since AdS curvature $R^{(d+1)}$ (\ref{4}) is not small. The simple resummation procedure and auxiliary mass method proposed in this paper permit to absorb curvature of the AdS background in zero order term of the expansion and give S-DW expansion of effective action not in powers of $R^{(d+1)}$ (symbolically) but in powers of derivatives of metric ${\tilde g}_{\mu\nu}(x)$. This procedure works for any warped product manifold which is conformally direct product (like (\ref{2})). S-DW expansion on the warped product spaces $M_{1}\otimes M_{d}$ was considered e.g. in \cite{Kirsten}, however method proposed below proves to be essentially more simple and gives immediately calculable Schwinger-DeWitt (Gilkey-Seely) coefficients on the AdS background.

We define quantum effective action $W$ following the approach of paper \cite{Mitra} as a difference of two actions calculated for Green functions having 'regular' and 'irregular' asymptotics at the AdS horizon. The ratio of corresponding functional determinants does not contain bulk contributions, in particular it is not plagued by the bulk quantum divergencies. The issue of relation between functional determinants of the operators acting in the bulk and on the boundary was widely discussed in context of double-trace deformations in AdS/CFT, see e.g. \cite{Hartman}, \cite{Diaz}. General, applicable for fields of any spin $s$, method of calculation of ratio of functional determinants differing only by the boundary conditions of the eigenfunctions of one and the same $(d+1)$ dimensional differential operator is developed in \cite{Barv2014} (and earlier in \cite{Barv2005}) where it is shown that this ratio always reduces to functional determinant of certain boundary-to-boundary operator in $d$ dimensions (cf. item 5.2 in Conclusion).
 
It will be shown also that expression for induced quantum vacuum energy density received in \cite{Mitra} (and repeated in \cite{Hartman}, \cite{Diaz}) is incompatible with direct calculation of the boundary free vacuum expectation value (VEV) of the stress-energy tensor of scalar field ($\langle 0 | T_{A}^{B} | 0 \rangle \sim \delta_{A}^{B}$) obtained in \cite{Saharian} with use of Wightman function. Whereas the auxiliary mass method proposed below gives exactly the same dependence of quantum vacuum energy on mass $m$ of scalar field as in \cite{Saharian}.

S-DW expansion makes sense only if there is some inverse mass squared parameter of expansion. It is shown in the paper that higher terms of proposed S-DW expansions are divergent in the boundary-free case $L=\infty$ and prove to be finite for finite $L$ being proportional to $(L^{2})^{n}$ that is to $(M_{{\it SM}}^{2})^{-n}$.

Induced quantum vacuum energy density in 4 dimensions as a function of the IR boundary location $L$, or equivalently as a function of $M_{\it SM}$, resembles Coleman-Weinberg potential with two extremums: one at $M_{\it SM} = 0$ and the other non-trivial one at $M_{\it SM} \ne 0$ which violates conformal symmetry and hopefully provides IR brane stabilization and permits to obtain the observed small value of mass hierarchy (see Sec. 5.4).

The outline of the paper is as follows. In the next section the resummation of S-DW expansion of Green function on the AdS background is performed and expression (\ref{25}) for the S-DW expansion of one loop quantum effective action is received by the 'auxiliary mass' method. In section 3 boundary-free case $L=\infty$ is elaborated, expressions for induced vacuum energy density in 5 dimensions and induced Planck mass in 4 dimensions are obtained, and it is shown that their correspondence with values (\ref{6}) (for $d=4$) of the conventional approach permits to express $M_{(4+1)}$ in (\ref{5}) and $m$ in (\ref{7}) through the scale $k$ of AdS space (\ref{2}); also the compatibility of expressions for induced quantum vacuum energy density with independently determined VEV of stress-energy tensor is discussed. In section 4 S-DW expansion is built for the case of IR-cut, $L<\infty$, of the AdS space and it is shown that coefficients of higher terms of the S-DW expansion are proportional to powers of $L$ and are expressed in an elementary way through coefficients of series of cylindrical functions at small argument. In Conclusion the possible links of results of the paper with other approaches, as well as possibility to fix mass hierarchy, are discussed. Also The Principle of Quantum Self-Consistency of the theory which unifies symbolically Sakharov's approach and bootstrap's 'no elementary particles/fields' basic idea is presented in Conclusion for reflection and criticism.

\section{S-DW expansion of the one loop action on AdS background: auxiliary mass method}

\qquad \subsection{\bf{Definition of $W[{\tilde g}_{\mu\nu}]$.}}

\quad Following \cite{Mitra} we define one loop quantum effective action $W$ as a difference of two actions calculated with use of scalar field Green functions $G_{\pm \nu}$ having 'regular' (+) and 'irregular' (-) asymptotic at the AdS horizon (see (\ref{11}) below):

\begin{equation}
\label{8}
W[{\tilde g}_{\mu\nu}] = \kappa (W_{+} - W_{-}) = - \frac{\kappa}{2} \,\ln \frac{{\rm Det}\hat D_{+}}{{\rm Det}\hat D_{-}},
\end{equation}
where $\kappa = \pm 1$. The virtue of this definition of one loop $W$ is absence in it of the bulk UV divergencies, as it was noted in the Introduction. Nevertheless this choice of $W$ is evidently arbitrary and only studying its physical consequences will permit to judge to what extent it is grounded and useful. In particular 'sign constant' $\kappa$ which we introduced in the r.h.s. of (\ref{8}) reflects the freedom of choice of the overall sign in definition of $W$. We must have in mind this arbitrariness when final results of calculations are analyzed.  

Differential operator $\hat D$ in (\ref{8}) is given by second variation over $\Phi$ of scalar field action (\ref{7}):

\begin{equation}
\label{9}
{\hat D} \Phi = \left[ -k^{2}z^{2}\frac{\partial^{2}}{\partial z^{2}} + (d-1) k^{2} z \frac{\partial}{\partial z} + m^{2} + k^{2}z^{2}(- {\tilde \triangle} + \mu^{2}) \right] \Phi = 0, 
\end{equation}
where $\tilde \triangle$ is d'Alembertian in $d$-space built with metric ${\tilde g}_{\mu\nu}(x)$, and looking ahead we introduced 'by hand' the auxiliary mass term $\mu^{2}$ which is absent in action (\ref{7}). In this paper introduction of auxiliary mass $\mu^{2}$ is a pure formal technique, perhaps it may be given some geometry meaning but we'll not speculate now about it.

Green functions corresponding to (\ref{9}) (with account of (\ref{2})) obey equation

\begin{equation}
\label{10}
{\hat D} G_{\pm\nu} (z,x; z',x') = (kz)^{d+1}\,\,\delta(z-z') \,\frac{\delta^{(d)} (x-x')}{\sqrt{\tilde g}}
\end{equation}
and following asymptotic conditions at the AdS horizon:

\begin{equation}
\label{11}
G_{\pm \nu} \to z^{\frac {d}{2} \pm \nu}\,\,\, z,\, z' \to 0,
\end{equation}
where

\begin{equation}
\label{12}
\nu = + \sqrt{\frac{d^{2}}{4} + \frac{m^{2}}{k^{2}}}.
\end{equation}

Boundary conditions 'on the other side' of the AdS space will be specified in sections 3, 4.
Following \cite{Witten}, \cite{Mitra} we'll consider as physically admissable the range $0<\nu^{2}<1$; in section 4 it will be shown that proposed 'auxiliary mass' method of calculation of coefficients of the S-DW expansion makes sense also only in this range of values of $\nu^{2}$. According to (\ref{12}) this range of values of $\nu$ is received for negative $m^{2}$. However there is another option: non-minimal scalar field action may be considered when action (\ref{7}) includes term $\xi\,R^{(d+1)}\Phi^{2}$. In this case expression (\ref{12}) for $\nu$ is modified in a well known way:

$$
\nu^{2}=\frac{1}{4}+\frac{m^{2}}{k^{2}}-(\xi - \xi_{c})d(d+1).  \eqno (12a)
$$
In particular for scale invariant in $(d+1)$ dimensions non-minimal scalar field action when $m=0$ and $\xi=\xi_{c}=(d-1)/4d$ one obtains in any dimension value $\nu = 1/2$ which falls into the admissable range $\nu^{2}<1$. However the case of non-minimal scalar field action has its own complications and in this paper we'll consider minimal action (\ref{7}) (some comments on non-minimal action see also in Sec. 3.1).

To calculate action $W$ (\ref{8}) we use the standard trick of variation of $W$ over any parameter '$\alpha$' of differential operator $\hat D$ (\ref{9}) with subsequent integration over this parameter:

\begin{eqnarray}
\label{13}
W[{\tilde  g}_{\mu\nu}]=\int_{\alpha_{0}}^{\alpha}\,d{\tilde\alpha}\frac{\partial W}{\partial {\tilde\alpha}} = {\rm{Tr}} \left\{ - \frac{\kappa}{2} \int_{\alpha_{0}}^{\alpha}\,d{\tilde\alpha} \left[{\bar G}_{\nu}\,\frac{\partial {\hat D}}{\partial{\tilde\alpha}}\right] \right\}, \nonumber \\
\\
{\bar G}_{\nu} \equiv G_{+\nu}-G_{-\nu}. \nonumber  \qquad \qquad \qquad \qquad
\end{eqnarray}

In our case, when only single scalar field is considered, ${\rm Tr}\,(...)$ is just $\int d\,{\it vol}_{(d+1)}$ (definition of $\int d\,{\it vol}_{(d+1)}$ is given in (\ref{5})). Difference of Green functions $\bar G_{\nu}$ evidently obeys homogeneous Eq. (\ref9); in the r.h.s. of (\ref{13}) it is taken at coinciding arguments: $z=z', x^{\mu}=x'^{\mu}$. Thus to get S-DW expansion of $W$ in powers of derivatives of metric ${\tilde g}_{\mu\nu}$ it is sufficient to obtain this expansion for ${\bar G}_{\nu}$ in (\ref{13}).

The conventional choice of parameter in (\ref{13}) is $\alpha = m^{2}$ \cite{Birrell}, \cite{Mitra}, etc. We'll see however that for the goal of building the S-DW expansion on the AdS background the choice of auxiliary mass $\mu^{2}$ in (\ref{9}) as a parameter $\alpha$ in (\ref{13}) is essentially more practical.

\qquad \subsection{\bf{S-DW resummation on AdS background}}

\qquad It will be shown now that necessary resummation of the S-DW expansion is possible thanks to conformally direct product nature of the warped product metric (\ref{2}) where all dependence on coordinate $z$ is in conformal factor $a^{2}(z)= 1/k^{2}z^{2}$. Actually the procedure of resummation of the S-DW expansion proposed below may be applied for any manifold of type $ds^{2}=a^{2}(z_{i})[g_{ij}(z)dz^{i}dz^{j}+{\tilde g}_{\mu\nu}(x)dx^{\mu}dx^{\nu}]$.

Let us transform field $\Phi(z,x)$ in action (\ref{7}) and in Eq. (\ref{9}) in a way:

\begin{equation}
\label{14}
\Phi (z,x) = (kz)^{(d-1)/2}\, \varphi(z,x).
\end{equation}
\\
This gives instead of (\ref{7}), (\ref{9}) (with account of expression (\ref{12}) for $\nu$):

\begin{equation}
\label{15}
S_{{\it matter}} = - \frac{1}{2} \int \left[ \varphi_{,z}^{2} + \frac{1}{z^{2}}\left(\nu^{2} - \frac{1}{4}\right) \varphi^{2} + {\tilde g}^{\mu\nu}\varphi_{,\mu}\varphi_{,\nu}\right]\,dz\,\sqrt{\tilde g}\,d^{d}x,
\end{equation}

\begin{equation}
\label{16}
{\bf\hat D}\,\varphi = \left\{\left[-\frac{\partial^{2}}{\partial z^{2}} + \frac{1}{z^{2}}\left(\nu^{2}-\frac{1}{4}\right)\right] + \left[-{\tilde \triangle} + \mu^{2}\right]\right\} \varphi \equiv ({\bf\hat D}_{z}+{\bf\hat D}_{x})\,\varphi = 0.
\end{equation}
Here ${\bf\hat D}_{z}$, ${\bf\hat D}_{x}$ symbolize corresponding terms in square brackets in (\ref{16}).
Presence in ${\bf\hat D}_{x}$ of auxiliary 'mass' term $\mu^{2}$ means in practice that in all subsequent expressions for Green functions in momentum space in flat $d$ dimensions it is necessary to substitute:

\begin{equation}
\label{17}
p \to \omega = \sqrt{p^{2}+\mu^{2}} \qquad \left(p \leftrightarrow \sqrt{-{\tilde \triangle}^{(0)}},\,\, \omega \leftrightarrow \sqrt{-{\tilde \triangle}^{(0)}+\mu^{2}}\right).
\end{equation}

Green function ${\bf G}(z,x;z'x')$ of differential operator ${\bf\hat D}$ (\ref{16}) is given by the equation:

\begin{equation}
\label{18}
{\bf\hat D}{\bf G}=\delta(z-z')\,\frac{\delta^{(d)}(x-x')}{\sqrt{\tilde g}},
\end{equation}
and according to (\ref{14}) it is connected with Green function $G$ determined by Eq. (\ref{10}) by the relation:

\begin{equation}
\label{19}
G(z,x;z',x')=(kz)^{(d-1)/2}\,{\bf G}(z,x;z'x')\,(kz')^{(d-1)/2}.
\end{equation}

Resummation of the Schwinger-DeWitt expansion of Green function ${\bf G}$ is possible since differential operators ${\bf\hat D}_{z}$ and ${\bf\hat D}_{x}$ in (\ref{16}) commute and heat 
kernel ${\rm exp}\, [-\tau {\bf\hat D}]$ may be factorized: ${\rm exp}\,[-\tau ({\bf\hat D}_{z}+{\bf\hat D}_{x})] ={\rm exp} \,[-\tau {\bf\hat D}_{z}]\,\cdot\, {\rm exp}\,[-\tau {\bf\hat D}_{x}]$. Because of it S-DW expansion of ${\bf G}$ at coinciding arguments may be received from (\ref{18}), (\ref{16}) in a standard way (expression for ${\bf\hat D}_{x}$ from (\ref{16}) is used in (\ref{20})):

\begin{eqnarray}
\label{20}
{\bf G}(z,x;z'x')|_{zx=z'x'}={\bf\hat D}^{-1}=\int_{0}^{\infty}\,d\tau \left\langle z,x \left| e^{-\tau ({\bf\hat D}_{z}+{\bf\hat D}_{x})}\right| z,x \right\rangle = \nonumber \\
\nonumber \\
=\int_{0}^{\infty}\,d\tau \left\langle z \left|e^{-\tau {\bf\hat D}_{z}} \right| z \right\rangle \cdot \left\langle x \left| e^{-\tau (-{\tilde\triangle}+\mu^{2})}\right| x \right\rangle = \qquad  \qquad  \nonumber \\
\nonumber \\
=\int_{0}^{\infty}\,d\tau \left\langle z \left|e^{-\tau {\bf\hat D}_{z}} \right| z \right\rangle \cdot \sum_{n=0}^{\infty}a_{n}(x,x)\tau^{n} \left\langle x \left| e^{-\tau(-{\tilde\triangle}^{(0)}+\mu^{2})}\right| x \right\rangle = \nonumber \\
\nonumber \\
=\sum_{n=0}^{\infty}a_{n}(x,x)\left(-\frac{\partial}{\partial\mu^{2}}\right)^{n} \int_{0}^{\infty}\,d\tau \left\langle z,x \left|e^{-\tau [ {\bf\hat D}_{z}+({-\tilde\triangle}^{(0)}+\mu^{2})]}\right| z,x \right\rangle = \nonumber
\\
=\sum_{n=0}^{\infty}a_{n}(x,x)\left(-\frac{\partial}{\partial\mu^{2}}\right)^{n}\,{\bf G}^{(0)}(z,x;z'x'|\mu^{2})|_{zx=z'x'}, \qquad \qquad
\end{eqnarray}
where $\mu^{2}$ must be put to zero after all differentiations over $\mu^{2}$ are fulfilled; ${\tilde\triangle}^{(0)}$ is d'Alembertian on the flat $d$-space ${\tilde g}_{\mu\nu}=\delta_{\mu\nu}$; ${\bf G}^{(0)}(z,x;z'x')$ in the last line obeys Eq. (\ref{18}) taken on the flat $d$-space.

Schwinger-DeWitt (Gilkey-Seely) coefficients $a_{n}(x,x)$ are well known for the heat kernel of d'Alembertian ${\tilde\triangle}[{\tilde g}_{\mu\nu}(x)]$.
First two of them are:

\begin{equation}
\label{21}
a_{0}(x,x)=1, \qquad a_{1}(x,x)=\frac{1}{6}{\tilde R}^{(d)},
\end{equation}
coefficient $a_{2}(x,x)$ contains terms $\sim{\tilde R}^{(d)2}$ (symbolically) and may include the gauge field Lagrangian - in case field $\Phi$ is charged.

Thus the last line in (\ref{20}) presents the looked for resummation of the S-DW expansion of Green function ${\bf G}$.
Since Green function $G$ is simply expressed through the Green function ${\bf G}$ (see  (\ref{19})) the S-DW expansion (\ref{20}) of ${\bf G}$ gives the needed expansion for $G$. Hence for difference $(G_{+\nu}-G_{-\nu}) = {\bar G}_{\nu}$ at coinciding arguments it is obtained:

\begin{equation}
\label{22}
{\bar G}_{\nu}[{\tilde g}_{\mu\nu}]_{z,x=z',x'}=\sum_{0}^{\infty}\,a_{n}(x,x)\,\left(-\frac{\partial}{\partial \mu^{2}}\right)^{n}\,F(z,\mu^{2})|_{\mu^{2}=0},
\end{equation}
where

\begin{equation}
\label{23}
F(z,\mu^{2})\equiv {\bar G}_{\nu}^{(0)}|_{z,x=z',x'}=\int\,\frac{d^{d}p}{(2\pi)^{d}}\,[G_{+\nu}^{(0)}(z,z,p,\mu)-G_{-\nu}^{(0)}(z,z,p,\mu)],
\end{equation}
\\
and $G_{\pm\nu}^{(0)}(z,z',p,\mu)$ are solutions of the equation (see (\ref{12}), (\ref{17}) for $\nu$, $\omega$):

\begin{eqnarray}
\label{24}
\left[-z^{2}\frac{\partial^{2}}{\partial z^{2}}+(d-1)z\frac{\partial}{\partial z}+\left(\nu^{2}-\frac{d^{2}}{4}\right)+z^{2}\omega^{2}\right]\,G_{\pm\nu}^{(0)}(z,z',p,\mu)= \\ \nonumber
= \frac{(kz)^{d+1}}{k^{2}}\,\delta(z-z'),
\end{eqnarray}
\\
satisfying asymptotic boundary conditions (\ref{11}). Boundary conditions at large $z=L$ (see (\ref{3})) will be specified in sections 3 and 4.

Using S-DW expansion of Green functions (\ref{22}) the corresponding S-DW expansion of the one loop quantum effective action $W$ may be obtained from (\ref{13}) where auxiliary mass $\mu^{2}$ in differential operator (\ref{9}) is taken as parameter $\alpha$. We also put in (\ref{13}) $\alpha_{0}\equiv \mu_{0}^{2}=\infty$ and define $W[{\tilde g}_{\mu\nu}]$ at zero value of the auxiliary mass, i.e. at $\alpha \equiv \mu^{2}=0$. Then the following expression for the one loop effective action is finally obtained ($F(z,\mu^{2})$ is defined in (\ref{23}) and relation $\partial{\hat D}/\partial \mu^{2}=k^{2}z^{2}$ from (\ref{9}) is used here):

\begin{eqnarray}
\label{25}
W[{\tilde g}_{\mu\nu}] = \int\,d\,{\it vol}_{(d+1)}\left\{\left(\frac{\kappa}{2}\right)\int_{0}^{\infty}\,d\mu^{2}\,{\bar G}_{\nu}|_{zx=z'x'}\cdot(k^{2}z^{2})\right\}= \qquad \nonumber
\\
= \int\,\, d{\it vol}_{(d+1)} [L_{(0)}+L_{(1)}+L_{(2)}+...] = \qquad  \qquad  \qquad \qquad \nonumber \\
\\
=\int\,d^{d}x\,\sqrt{{\tilde g}}\,\int_{\epsilon}^{L}\frac{dz}{(kz)^{d+1}}\,\left[\left(\kappa\,\frac{k^{2}z^{2}}{2}\right)\int_{0}^{\infty}\,d\mu^{2}\,F(z,\mu^{2}) + \right. \qquad \qquad \nonumber \\
\nonumber \\
+ \left.\kappa\,\frac{k^{2}z^{2}}{2} \,\frac{1}{6} \,  F(z,0) \, {\tilde R}^{(d)} + \kappa\,\frac{k^{2}z^{2}}{2}\,\sum_{n=1}^{\infty}\,a_{n+1}(x,x)\left(-\frac{\partial}{\partial \mu^{2}}\right)^{n}\,F(z,\mu^{2})\,|_{\mu^{2}=0}\right] \nonumber
\end{eqnarray}

First two terms in square brackets in (\ref{25}) respectively are: induced vacuum energy density and induced Einstein term. Every differentiation over $\mu^{2}$ in (\ref{25}) gives additional multiplication by $z^{2}$ (see (\ref{27}) below), thus expansion (\ref{25}) in a number of its first terms depending on dimensionality $d$ is, as it should be, an expansion in decreasing powers of divergencies in integral over $z$ at its lower limit (in case $\epsilon \to 0$). 

We'll see in section 3.3 that in the boundary free case $L=\infty$ first derivative $\partial F(z,\mu^{2})/\partial \mu^{2}$ in (\ref{25}) at $\mu^{2}=0$ is finite, whereas all others are divergent at $\mu^{2}=0$. This is a manifestation of absence of small $(mass)^{-2}$ parameter of S-DW expansion (\ref{25}) when $L=\infty$. It will be shown in section 4 that introduction of the 'visible' brane at $z=L$ will give S-DW expansion (\ref{25}) in powers (symbolically) $({\tilde R}^{(d)}L^{2})^{n} = ({\tilde R}^{(d)}/M_{\it SM}^{2})^{n}$. 

\section{S-DW expansion on AdS in the boundary-free case}

\qquad \subsection{Induced vacuum energy density and Planck mass}

\quad To receive S-DW expansion (\ref{25}) of the one loop quantum action in the boundary-free case, $L=\infty$, the conventional condition $G_{\pm\nu}^{(0)} \to 0$ at $z,z' \to \infty$ must be imposed on Green functions plus to asymptotic conditions (\ref{11}) at the horizon. Solutions of Eq. (\ref{24}) satisfying these boundary conditions are well known:

\begin{equation}
\label{26}
G_{\pm\nu}^{(0)}(z,z',p,\mu)=(k^{2}zz')^{d/2}\,k^{-1}\,[I_{\pm\nu}(\omega z)K_{\nu}(\omega z')\theta(z'-z)+ (z \leftrightarrow z')],
\end{equation}
$I_{\pm\nu}, \,K_{\nu}$ are cylindrical functions of imaginary argument. For difference (\ref{23}) of Green functions (\ref{26}) at coinciding arguments it is immediately  obtained (cf. \cite{Mitra} in case $\mu = 0$, i.e. for $\omega = p$; subscript $\infty$ means that $F_{\infty}$ is calculated at $L=\infty$):

\begin{eqnarray}
\label{27}
F_{\infty}(z,\mu^{2})=(kz)^{d}k^{-1}\left(-\frac{2\sin(\pi\nu)}{\pi}\right)\int\,\frac{d^{d}p}{(2\pi)^{d}}\,K_{\nu}^{2}(\omega z) = \qquad  \nonumber \\
\nonumber \\
=  - C_{d}\,\int_{\mu z}^{\infty}\,(y^{2}-\mu^{2}z^{2})^{\frac{d}{2}-1}\,K_{\nu}^{2}(y)\,y\,dy, \qquad  C_{d}= \frac{2\sin{\pi\nu}}{\pi}\,\frac{k^{d-1}\Omega_{d-1}}{(2\pi)^{d}},
\end{eqnarray}
where $y=\omega z$, $\Omega_{d-1}$ is volume of $(d-1)$ dimensional sphere. It is seen that condition of validity of (\ref{25}) $F_{\infty}(z,\mu^{2}) \to 0$ at $\mu^{2} \to \infty$ is fulfilled since $K_{\nu}^{2}(y)$ exponentially decreases at large argument and $y > \mu z$ in integral in (\ref{27}).

Differentiations of $F_{\infty}(z,\mu^{2})$ over $\mu^{2}$ in the r.h.s. of (\ref{27}) are elementary, results of integrations over $y$ (after $\mu^{2}$ is put equal to 0) are given by well known formula (8.339.2 in \cite{Ryzhik}):

\begin{equation}
\label{28}
\int_{0}^{\infty}\,y^{\lambda - 1}\, K_{\nu}^{2}(y)\,dy = \frac{2^{\lambda - 3}}{\Gamma (\lambda)}\,\Gamma^{2}(\frac{\lambda}{2})\,\Gamma(\frac{\lambda}{2}+\nu)\,\Gamma(\frac{\lambda}{2}-\nu),
\end{equation}
$\Gamma (x)$ is Euler Gamma function.

Now let us consider 5-dimensional RS-model, i.e. $d=4$. With substitution of $F_{\infty}(z, \mu^{2})$ from (\ref{27}) into the first term of S-DW expansion (\ref{25}) the following expression for quantum induced vacuum energy density in 5 dimensions is obtained ($u\equiv \mu z$, $C_{4}$ is defined in (\ref{27})):

\begin{eqnarray}
\label{29}
V^{(5)}_{\it vac}=-L^{(0)}=\kappa\,k^{2}\,C_{4}\,\int_{0}^{\infty}\,u\,du\,\int_{u}^{\infty}\,(y^{2}-u^{2})\,K_{\nu}^{2}(y)\,y\,dy= \qquad \nonumber \\
\nonumber \\
\qquad = \frac{\kappa\, k^{2}\,C_{4}}{4}\,\int_{0}^{\infty}\,y^{5}K_{\nu}^{2}(y)\,dy= \kappa\,\frac{k^{5}}{60\pi^{2}}\,(4-\nu^{2})\,(1-\nu^{2})\,\nu. \qquad  \qquad 
\end{eqnarray}
As it is physically expected $V^{(5)}_{\it vac}=0$ at $m^{2}=k^{2}(\nu^{2}-4)=0$ (see (\ref{12}) for $d=4$).

Induced Einstein action in 4 dimensions is given by the second term in S-DW expansion (\ref{25}). $F_{\infty}(z, 0)$ is calculated from (\ref{27}), (\ref{28}) and is given by expression \cite{Mitra}: 

\begin{equation}
\label{30}
F_{\infty}(z, 0)=\left[G_{+\nu}^{(0)}-G_{-\nu}^{(0)}\right]_{z,x=z',x'} = - \frac{k^{3}}{12\pi^{2}}(1-\nu^{2})\nu;
\end{equation}

Thus for induced Einstein term it is obtained from (\ref{25}):

\begin{equation}
\label{31}
M^{2}_{\it Pl}\,{\tilde R}^{(4)}= -\frac{\kappa}{48\pi^{2}\,\epsilon^{2}}\, (1-\nu^{2})\,\nu \, \frac{1}{6}\cdot {\tilde R}^{(4)},
\end{equation}

According to our definition (\ref{12}) $\nu > 0$, and permitted range of values of $\nu$ is $0 <\nu^{2} <1$. Thus it is seen from (\ref{29}), (\ref{31}) that choice $\kappa = -1$ in definition of $W$ (\ref{8}) gives physically acceptable $M_{\it Pl}^{2} > 0$, $V_{\it vac}^{(5)} < 0$.

Let us map values of induced vacuum energy density (\ref{29}) and induced Planck mass in (\ref{31}) with expressions for bulk vacuum energy density in 5 dimensions and Planck mass in 4 dimensions in Randall-Sundrum model (expressions (\ref{6}) for $d=4$):

\begin{equation}
\label{32}
V^{(5)}_{\it{vac\,(RS)}} = - 12\,k^{2}M_{(5)}^{3}, \,\,\, \qquad M^{2}_{\it {Pl\,(RS)}} = \frac{M_{(5)}^{3}}{2\,k^{3}\epsilon^{2}}
\end{equation}

Naive equating quantum induced values (\ref{29}), (\ref{31}) (where the choice $\kappa = -1$ is made) to the corresponding values (\ref{32}) of the RS-model determines $M_{(5)}$ through $k$ and also demands $\nu^{2}=-1$ (i.e. $m^{2}/k^{2}= -5$). This unphysical result should not be taken too seriously because it is affected by a set of precondi
tions, not to mention that there is lack of clear understanding of very notion of quantum self-consistency of the theory. 

We note that the sign of induced Planck mass (\ref{31}) may change to the opposite one if non-minimal scale-invariant scalar field action with additional term $\xi\,R^{(d+1)}\Phi^{2}$ in (\ref{7}) is considered instead of minimal action (\ref{7}). Many previous expressions as a functions of $\nu$ are valid for non-minimal action where value of $\nu$ is given now by (12a). For $\xi \ne 0$ asymptotic expansion coefficients will acquire additional terms, in particular in coefficient $a_{1}$ in (\ref{21}) 1/6 must be changed to $(1/6 - \xi)$; the same substitution of $1/6 \to (1/6 - \xi)$ must be performed in expressions (\ref{25}), (\ref{31}). For conformally invariant scalar field action in 5 dimensions when $d=4$, $m = 0$, $\xi=\xi_{c}= 3/16$ and $\nu = 1/2$ we have $(1/6 - \xi_{c})= - 1/48 <0$. Thus induced $M^{2}_{\it Pl}$ in (\ref{31}) is positive in this case if  $\kappa = +1$ in (\ref{8}), (\ref{25}) and hence in (\ref{31}). Induced vacuum energy (\ref{29}) is positive in this case, however it is unclear if (\ref{29}) is valid at all in the non-minimal case. The one loop effective action of the non-minimal scalar field on the AdS background needs further research.

The idea to receive AdS solution of Einstein equations and even to obtain reasonable branes' stabilization mechanism in the RS-model by quantum generating the needed right hand side of the equations is quite popular. The next step would be to try to follow Sakharov and to induce quantumly the left hand side of Einstein equations. We don't know how to do it for Einstein equations in $(d+1)$ dimensions, but it was demonstrated above that self-consistency demand of coincidence of two values (\ref{29}), (\ref{31}) with those ones in (\ref{32}) determines otherwise arbitrary constants of the theory - $M_{(5)}$ and mass $m$ of scalar field. Perhaps this promising result may serve an inspiration for the search of some general Principle of Quantum Self-Consistency - see also comments in item 5.3 of Conclusion.

\qquad \subsection{Strange discrepancy in calculations of induced vacuum energy density}

\qquad In \cite{Mitra} (by Gubser and Mitra, see also \cite{Hartman}, \cite{Diaz}) one loop quantum vacuum energy density $V^{(5)}_{\it vac(GM)}$ was calculated from expression (\ref{30}) for difference of Green functions $G_{\pm\nu}$ using trick (\ref{13}) with a conventional choice $\alpha = m^{2}$. For $d=4$ it was received in \cite{Mitra}:

\begin{eqnarray}
\label{33}
V_{\it vac(GM)}^{(5)}=V_{+}-V_{-} = \frac{1}{2} \int_{m^{2}_{BF}}^{m^{2}}\,d{\tilde m}^{2}\,\left[G_{+\nu}^{(0)}-G_{-\nu}^{(0)}\right] = \nonumber \\
\nonumber \\
= - C_{4} \int_{0}^{\nu}k^{2}{\tilde\nu}\,d{\tilde\nu}\int_{0}^{\infty}K^{2}_{\nu}(y) \,y^{3}\,dy = - \frac{k^{5}}{12\pi^{2}} \left(\frac{\nu^{3}}{3} - \frac{\nu^{5}}{5}\right),
\end{eqnarray}
where $m_{BF}$ is Breitenlohner-Freedman mass corresponding to $\nu = 0$ in (\ref{12}) and expression (\ref{27}) for difference of Green functions (cf. (\ref{23}), (\ref{30})) was used for $\mu = 0$ and $d=4$.

Dependence of the right hand side of (\ref{33}) (the choice $\alpha = m^{2}$ in (\ref{13})) on $\nu$, that is on mass $m$ of scalar field, drastically differs from that in (\ref{29}) (the choice $\alpha = \mu^{2}$ in (\ref{13})). Where is the truth? To find an answer it is worthwhile to check up compatibility of expressions (\ref{29}) or (\ref{33})
with VEV of stress-tensor $T_{AB}$ of scalar field $\Phi$ received by variation of action (\ref{7}) over full metric $g_{AB}$ (\ref{2}) ($x^{A}=\{z, x^{\mu}\}$; $g$ is determinant of $g_{AB}$):

\begin{equation}
\label{34}
T_{AB} = - \frac{2}{\sqrt g} \frac{\delta S}{\delta g_{AB}} = \Phi_{,A}\Phi_{,B} - \frac{1}{2}g_{AB}[g^{MN}\Phi_{,M}\Phi_{,N} + m^{2}\Phi^{2}].
\end{equation}

Above named compatibility means that directly calculated VEV of $T_{AB}$ (\ref{34}) must coincide with the variation of zero order term of effective action $W_{(0)}$ over metric $g_{AB}$:

\begin{equation}
\label{35}
\langle 0 |\, T_{AB} \,| 0 \rangle = - \frac{2}{\sqrt g} \frac{\delta W_{(0)}}{\delta g_{AB}} = - g_{AB} \,V_{\it vac}.
\end{equation}

Boundary-free VEV of $T_{AB}$ on the AdS background was calculated in \cite{Saharian} with use of Wightman function of scalar field $\langle 0 |\,\Phi\,\Phi\,| 0 \rangle$. Formula (3.24) of \cite{Saharian} taken for $d=4$ and $\nu$ arbitrary gives:

\begin{equation}
\label{36}
\langle 0 |\, T_{A}^{B} \,| 0 \rangle = \delta_{A}^{B}\, \frac{k^{3}m^{2}}{60\pi^{2}}\,(\nu^{2}-1)\,\nu,
\end{equation}
which with account of (\ref{12}) coincides with expression in (\ref{29}) (where $\kappa = -1$). 

Thus for the choice $\kappa = -1$ in definition (\ref{8}) of $W$ (and hence in (\ref{29})) compatibility condition (\ref{35}) is fulfilled for $V^{(5)}_{\it vac}$ (\ref{29}). 

And compatibility condition (\ref{35}) is evidently not fulfilled for $V^{(5)}_{\it vac(GM)}$ (\ref{33}) considered in \cite{Mitra}, \cite{Hartman}, \cite{Diaz}. This is even more surprising since expression (2.39) in \cite{Saharian} for Wightman function used in \cite{Saharian} for calculation of VEV of $T_{AB}$ exactly coincides with expression (\ref{30}) for difference of '$\pm\nu$' Green functions used in \cite{Mitra} [formula (19) of \cite{Mitra}]. D.E. Diaz \cite{Diaz2} supposed that this mismatch is probably due to a Casimir-like missing contribution and paid attention to discussion about the "correct" stress-tensor in \cite{Herzog}, \cite{Lee}\footnote{The author is grateful to Danilo Diaz for these commentaries}.

In any case the question remains why in the proposed in the paper 'auxiliary mass' approach zero order 'vacuum energy' term (\ref{29}) of the S-DW expansion of effective action $W$ exactly coincides with the result of direct calculation of VEV of stress-energy tensor (\ref{34})? 

\subsection{Impossibility of S-DW expansion on the boundary free AdS}

\qquad The fourth and subsequent terms of S-DW expansion (\ref{25}) are divergent in the boundary-free case $L=\infty$ when in (\ref{25}) $F(z,\mu^{2}) = F_{\infty}(z, \mu^{2})$ (\ref{27}). We again consider $d=4$. First derivative of $F_{\infty}$ over $\mu^{2}$ at $\mu^{2}=0$ is finite: $(\partial F_{\infty}/\partial \mu^{2})_{\mu = 0}=k^{3}z^{2}\nu /8\pi^{2}$. Thus for the third term in (\ref{25}) after integration over $z$ it is obtained:

\begin{equation}
\label{37}
\int_{\epsilon}^{L}\,\frac{dz}{(kz)^{5}}\,L^{(2)}= - \frac{\kappa\,C_{4}}{2\,k^{3}}\,\ln\frac{L}{\epsilon}\,\int_{0}^{\infty}\,y\,K_{\nu}^{2}\,dy\cdot a_{2}= -\frac{\kappa\,\nu}{16\pi^{2}}\,\ln\frac{L}{\epsilon}\cdot a_{2} \quad.
\end{equation}

However higher derivatives of $F_{\infty}$ over $\mu^{2}$ at $\mu^{2}=0$ are not so well behaved. In particular for second derivative it is obtained from (\ref{27}):

\begin{equation}
\label{38}
\left(\frac{\partial}{\partial \mu^{2}}\right)^{2}\,F_{\infty}(z,\mu^{2})= -\frac{\sin\pi\nu}{\pi}\,\frac{k^{3}z^{4}}{8\pi^{2}}\,K_{\nu}^{2}(\mu z) \sim (\mu z)^{-2\nu} \to \infty \quad at \quad \mu \to 0.
\end{equation}
It is evident that all higher derivatives of $F_{\infty}(z,\mu^{2})$ (\ref{27}) over $\mu^{2}$ will be divergent at $\mu=0$ as well. In the next section we'll show that introduction of IR cut of AdS space at some $z=L$ makes expansion (\ref{25}) sensible.

\section{IR cut of AdS as a small parameter of \\
S-DW expansion}

\qquad Let us introduce IR cut of AdS space (\ref{2}) at $z=L$ (see (\ref{3})) and write down '$d$-space momentum' components of Green functions $G^{(0)}_{\pm\nu(L)}(z,z',p,\mu)$ satisfying Eq. (\ref{24}), asymptotic conditions (\ref{11}) at the AdS horizon, and one and the same Robin boundary condition at $z=L$:

\begin{equation}
\label{39}
\left[z\,\frac{\partial G^{(0)}_{\pm\nu (L)}(z,z',p,\mu)}{\partial z}+ r\, G^{(0)}_{\pm\nu (L)}(z,z',p,\mu)\right]_{z=L} = 0
\end{equation}
(the same at $z'=L$, $r$ is arbitrary constant).

Then, instead of expression (\ref{26}) which is valid for $L=\infty$, for these Green functions one obtains:

\begin{equation}
\label{40}
G^{(0)}_{\pm\nu (L)}=-\frac{(zz'k^{2})^{d/2}\,\pi}{k\,2\sin\pi\nu}\,\gamma^{\pm 1/2}(\omega L)\,[I_{\pm\nu}(\omega z)\,U_{\nu}(\omega z')\,\theta(z'-z) + (z \leftrightarrow z')\,],
\end{equation}
\\
where $U_{\nu}(\omega z)$ is a solution of homogeneous Eq. (\ref{24}) satisfying boundary conditions (\ref{39}) $(z\,U_{,z} + r\,U)_{z=L} = 0$:

\begin{eqnarray}
\label{41}
U_{\nu}(\omega z) = \gamma^{1/2}(\omega L)\,I_{\nu}(\omega z) - \gamma^{-1/2}(\omega L)\,I_{-\nu}(\omega z), \nonumber \\
\\
\gamma (\omega L) = \frac{\left(\frac{d}{2}+ r\right)\,I_{-\nu}(\omega L)+\omega L\,\,\frac{d\,I_{-\nu}(\omega L)}{d\,\omega L}}{\left(\frac{d}{2}+ r\right)\,I_{\nu}(\omega L)+\omega L\,\,\frac{d\,I_{\nu}(\omega L)}{d\,\omega L}} \qquad \nonumber.
\end{eqnarray}

For difference (\ref{23}) of these Green functions at coinciding arguments determining S-DW expansion (\ref{25}) it is obtained (instead of (\ref{27}) when $L=\infty$):

\begin{equation}
\label{42}
F_{L}(z,\mu^{2})= - C_{d}\,\int_{\mu z}^{\infty}\,(y^{2}-\mu^{2}z^{2})^{\frac{d}{2}-1}\,E(y,yL/z)\,y\,dy,
\end{equation}
where $C_{d}$ is defined in (\ref{27}), $y=\omega z$, $yL/z = \omega L$, $\omega = \sqrt{p^{2}+\mu^{2}}$,

\begin{equation}
\label{43}
E(y, yL/z)= \left(\frac{\pi}{2\sin\pi\nu}\right)^{2}\,[\gamma\,I_{\nu}^{2}(y)+\gamma^{-1}\,I_{-\nu}^{2}(y) - 2I_{\nu}(y)I_{-\nu}(y)],
\end{equation}

It is immediately seen from (\ref{41}) that $\gamma \to 1$ exponentially at $L \to \infty$ , hence $E \to K_{\nu}^{2}$, i.e. expression (\ref{42}) for $F_{L}(z,\mu^{2})$ reduces to (\ref{27}) at $L \to \infty$ as it could be expected.

The S-DW expansion of $W$ on the AdS space with IR boundary is obtained if $F_{L}(z,\mu^{2})$ is used in (\ref{25}). $F_{L}$ (\ref{42}) is received from $F_{\infty}$) by substitution in (\ref{27}) instead of $K_{\nu}^{2}(y) expression E(y, yL/z)$ given in (\ref{43}). Since $L \gg (k^{-1},\,\epsilon)$ this will not affect essentially three first terms $a_{0}, a_{1}, a_{2}$ (divergent at $\epsilon \to 0$ for $d=4$) of the S-DW expansion calculated for $L=\infty$. That is r.h.s. of expressions for vacuum energy density (\ref{29}) and Planck mass (\ref{31}) (as well as (\ref{37})) are only slightly corrected in case $L < \infty$ (however dependence of the vacuum energy density on $L$ proves to be non-trivial - see item 5.4 in Conclusion).

Most interesting is comparison of higher terms of S-DW expansion (\ref{25}) calculated for $L=\infty$ and for $L$ finite. In case $L=\infty$ second derivative of $F_{\infty}(z,\mu^{2})$ (\ref{27}) over $\mu^{2}$ is divergent at $\mu = 0$ - see (\ref{38}). Whereas second derivative of $F_{L}(z,\mu^{2})$ (\ref{42}) is well defined at $\mu = 0$. Expression for it (for $d=4$) is like (\ref{38}) where again $K_{\nu}^{2}(\mu z)$ must be replaced by $E(\mu z, \mu L)$ (\ref{43}):

\begin{eqnarray}
\label{44}
\left(\frac{\partial}{\partial \mu^{2}}\right)^{2}\,F_{L}(z,\mu^{2})= -\frac{\sin\pi\nu}{\pi}\,\frac{k^{3}z^{4}}{8\pi^{2}}\,E(\mu z, \mu L)= \qquad  \qquad  \nonumber \\
\nonumber \\
= - \frac{\pi}{\sin\pi\nu}\,\frac{k^{3}z^{4}}{32\,\pi^{2}} [\gamma (\mu L)\, I_{\nu}^{2}(\mu z) + \gamma^{-1}(\mu L)\,I_{-\nu}^{2}(\mu z) - 2\,I_{\nu}(\mu z)\,I_{-\nu}(\mu z)] = \\
\nonumber \\
= -\frac{k^{3}z^{4}}{32\pi^{2}\,\nu}\,\left\{\left[c\,\left(\frac{z}{L}\right)^{2\nu}+\frac{1}{c}\,\left(\frac{L}{z}\right)^{2\nu}-2\right] + \mu^{2}L^{2}\,Q_{1}+\mu^{4}L^{4}\,Q_{2}+\ldots\right\}, \nonumber \\
\nonumber
\end{eqnarray}
where 
$$
c=\frac{d+r-\nu}{d+r+\nu} \qquad here \,\, \, d=4, \qquad \qquad 
$$
$\gamma (\mu L)$ is defined in (\ref{41}); $Q_{1}$, $Q_{2}$... have the same structure as the first term in square brackets in the last line of (\ref{44}) with the difference that in $Q_{1}, Q_{2}\ldots$ each of three terms in square brackets is provided with factors - polynomials in $\alpha^{2}\equiv (z/L)^{2}$: $Q_{1}=a_{1}\alpha^{2}+b_{1}$, $Q_{2}=a_{2}\alpha^{4}+b_{2}\alpha^{2}+c_{2}$, etc., which coefficients $a_{i}, b_{i}, c_{i}\ldots$ are functions of $\nu, d, r$ and are easily calculable from the elementary expansions of Bessel functions at small argument.

Let us look (again for $d=4$) at the fourth term ($a_{3}$) of expansion (\ref{25}) determined by the second derivative of $F_{L}(z,\mu^{2})$ over $\mu^{2}$ at $\mu=0$ that is determined by the r.h.s. of (\ref{44}) at $\mu=0$:

\begin{equation}
\label{45}
\int_{\epsilon}^{L}\,\frac{dz}{(kz)^{5}}\,L_{(3)}=\frac{\kappa}{64\pi^{2}\nu}\,\int_{\epsilon / L}^{1}\,\alpha\,d\alpha\,[c\,\alpha^{2\nu}+c^{-1}\,\alpha^{-2\nu} - 2] \cdot L^{2}\,a_{3}, 
\end{equation}
$\alpha = z/L$. Divergence of integral in (\ref{45}) in UV, i.e. at $\epsilon = 0$, would undermine the very idea of expansion (\ref{25}). Most dangerous in this respect is second term in the r.h.s. of (\ref{45}) $\int\,\alpha^{1-2\nu}\,d\alpha$ which is finite at $\epsilon = 0$ when $\nu < 1$ i.e. in the range $0<\nu^{2}<1$ considered in \cite{Witten}, \cite{Mitra}. The same condition guarantees finiteness at $\epsilon = 0$ of all higher terms of expansion (\ref{25}). These terms are proportional to $L^{2n - 4}$: $L^{2}a_{3}$, $L^{4}a_{4}\ldots$, $L^{2n-4}a_{n}\ldots$. 

Thus S-DW expansion (\ref{31}) is really expansion in $L^{2}$ and makes sense if $L^{2}{\tilde R}^{(d)} = {\tilde R}^{(d)}/\,M_{SM}^{2} \ll 1$.

\section{Conclusion: four remarks for future work}

\subsection{Possible link with gravity polarization operator}

\qquad Let us forget now about auxiliary mass $\mu$ (i.e. put $\mu^{2}=0$ in Eq. (\ref{9}), etc.) and look at the variations over metric $g_{AB}$ of effective action $W$ defined in (\ref{8}). First variation will give VEV of energy-momentum tensor (\ref{34}) expressed in a standard way through second derivatives $\nabla_{\mu}\nabla_{\nu'}$, $\nabla_{z}\nabla_{z'}$ at coinciding arguments of difference ${\bar G}_{\nu}$ of Green functions defined in (\ref{13}). These calculations were performed in \cite{Saharian} with use of Wightman function and are confirmed in Sec. 3 above (see expressions (\ref{35}), (\ref{36}) and (\ref{29})). 

Second variation of $W$ (\ref{8}) over metric gives gravity field polarization operator $\langle 0 | T_{AB}T_{CD} | 0 \rangle$ which is expressed now also in a standard way through derivatives of difference ${\bar \Pi}$ of two scalar field polarization operators built from Green functions $G_{\pm\nu} (z,x;z',x')$. In momentum $d$-space:

\begin{eqnarray}
\label{46}
{\bar \Pi}(z, z', p^{2}) = \Pi_{+\nu} - \Pi_{-\nu} = \qquad  \qquad  \qquad  \qquad \nonumber
\\
\\ \nonumber
\int \,\frac{d^{d}q}{(2\pi)^{d}}\left[G_{+\nu}(z,z',q)G_{+\nu}(z',z,p-q) - G_{-\nu}(z,z',q)G_{-\nu}(z',z,p-q)\right],
\end{eqnarray}
where $G_{\pm\nu}(z,z',q)$ are given by the $\mu^{2} = 0$ version of formulas (\ref{26}) - for $L=\infty$ or (\ref{40}) 
 - for $L < \infty$.

Integral over $q$ in (\ref{46}) is convergent in UV since at $q\to \infty$ expression in square brackets in (\ref{46}) $[\ldots ] \sim (I_{\nu}+I_{-\nu})\,K_{\nu}^{3} \sim e^{-2qz}$ (for $z=z'$). However in the boundary-free case $(L=\infty)$ integral in (\ref{46}) is IR divergent at small $q$ when higher derivatives of ${\bar \Pi}(z, z', p^{2})$ over $p^{2}$ are calculated at $p^{2}=0$. This resembles divergence of expression (\ref{38}) at $\mu^{2}=0$. Preliminary investigation shows that introduction of finite $L < \infty$ regularize these IR divergences and that Taylor series expansion of ${\bar \Pi}_{L}(z, z', p^{2})$ in $p^{2}$ is similar to expansion (\ref{44}) of second derivative of $F_{L}(z, \mu^{2})$ in $\mu^{2}$. This parallel needs further investigation.

\subsection{Bulk-boundary correspondence} 

\qquad It would be interesting to apply general approach of paper \cite{Barv2014} (formula (4.14) of \cite{Barv2014}) to calculation of effective action (\ref{8}). In this approach ratio of bulk determinants ${\rm Det}{\hat D}_{+}/{\rm Det}{\hat D}_{-}$ is expressed through the ratio of determinants ${\rm det} F_{+}/{\rm det} F_{-}$ of boundary-to-boundary operators 

$$
F_{\pm}= -G_{(D),zz'}|_{z,z'=\epsilon}+f_{\pm}. 
$$
In the context of \cite{Barv2014} boundary must be placed at $z=\epsilon$ (we take here $L=\infty$ like in Sec. 3). 

Non-local 'Robin coefficients' of generalized Neumann boundary conditions at $z=\epsilon$: $f_{\pm}= - \phi_{\pm,z}(z,p)/\phi_{\pm}(z,p)|_{z=\epsilon}$ are easily received from the expressions for solutions $\phi_{\pm}$ of homogeneous Eq. (\ref{24}) having asymptotics (\ref{11}) at the horizon $z=0$. 

$G_{(D)}$ is the Dirichlet Green function equal to zero at $z,z'=\epsilon$ which is certain superposition of Green functions $G_{\pm\nu}$ (\ref{26}). 

Summing it all gives nice result:

\begin{equation}
\label{47}
\frac{{\rm Det}{\hat D}_{+}}{{\rm Det}{\hat D}_{-}} = \frac{{\rm det} F_{+}}{{\rm det} F_{-}}=\Pi_{p}\,\frac{I_{-\nu}(|p|\epsilon)}{I_{+\nu}(|p|\epsilon)},
\end{equation}
where product over $d$-space momentum $p$ is finite since ratio $I_{-\nu}/I_{+\nu} \to 1$ exponentially at large argument. However final calculations prove to be more difficult in this approach then in the present paper where bulk formulas were considered. Thus we leave comparison of two approaches for the future work.

\subsection{Principle of Quantum Self-Consistency} 

\qquad It was demonstrated in Sec. 3.1 that naive equating of dynamical constants (vacuum energy density in 5 dimensions and Planck mass in 4 dimensions) of conventional RS-model to those ones which are induced from the one loop quantum matter action determines through AdS scale $k$ two $ad \,\, hoc$ constants of the model (Planck mass in 5 dimensions and mass $m$ of scalar field). And although in the simplified model considered in the paper $m$ proves to be unphysical  ($\nu^{2}=-1$ that is $m^{2}/k^{2}=-5$) the very opportunity to decrease number of arbitrary constants of the theory by imposing certain self-consistency conditions seems to be quite interesting. 

Plethora of $ad\,\,hoc$ 'fundamental' constants in Standard Model (not to mention the arbitrary values of fine-structure constant, mass hierarchy, etc.) is a 'disease' of theoretical physics during decades. Unfortunately neither  string theory and higher dimensions, nor QCD or brilliant idea of $S$-matrix bootstrap (no elementary particles, all particles are bound states of the same particles), nor modern promising AdS/CFT correspondence managed so far to cure this 'disease'.

In Sakharov's 'induced gravity' approach Planck mass is also a sort of secondary constant determined from quantum dynamics of the 'elementary' matter fields through the UV regularization parameter (in the present paper induced value of Planck mass is expressed through the location $z=\epsilon$ of the UV-cut of AdS space). Perhaps the combination of Sakharov's approach with the bootstrap basic idea of "no elementary particles/fields" will prove to be the looked for Quantum Self-Consistency Principle (QSCP) determining constants of the theory. The following "toy formula" for self-consistent effective action $W[\phi(j)]$ illustrates this idea (here $\phi (j)$ symbolizes all fields of any spin, including gravity; $j$ symbolizes collection of space-time coordinates and internal indexes): 

\begin{equation}
\label{48}
W[\phi] = \gamma\,{\rm ln}\,{\rm Det}\left[\frac{\delta^{2}W}{\delta \phi(j)\,\delta \phi(k)}\right],
\end{equation}
where $\gamma$ is some constant. Or perhaps in the right-hand-side of (\ref{48}) may be written the ratio of determinants like it was done in (\ref{8}) for effective action considered in this paper. In any case non-triviality of the quantumly self-induced action (\ref{48}) is immediately seen because free field theories described by quadratic action $W\sim \phi^{2}$ are evidently excluded by QSCP (\ref{48}).

In particular for pure gravity on the AdS background variation $h_{AB}$ of gravity field ($g_{AB} \to g_{AB}+h_{AB}$ where $g_{AB}$ is metric (\ref{2})) can be taken instead of quantum scalar field $\Phi$ in (\ref{7}). Then QSCP may look as (cf. (\ref{8})):

\begin{equation}
\label{49}
S[{\tilde g}_{\mu\nu}] = \gamma \,\ln \frac{{\rm Det}\hat S''_{+}[{\tilde g}_{\mu\nu}]}{{\rm Det}\hat S''_{-}[{\tilde g}_{\mu\nu}]},
\end{equation}
where $\hat S''[{\tilde g}_{\mu\nu}]$ is differential operator received by the second variation of action $S[g]$ over the metric on background (\ref{2}) or some its modifications. Naive equating in Sec. 3.1 of Planck masses (\ref{31}) and (\ref{32}) is an analogy of equating of Einstein terms in the l.h.s. and r.h.s. of QSCP (\ref{49}).

In conclusion we'll present rather curious solution of (\ref{48}) for the case of one-dimensional functional space when (\ref{48}) is just an ordinary differential equation $W(\phi)=\gamma\,{\rm ln}[d^{2}W/d\phi^{2}]$:

\begin{equation}
\label{50}
W[\phi] = \gamma\,{\rm ln}\,\left[1+{\rm tg}^{2}\left(\frac{\phi}{\sqrt {2\gamma}}\right)\right].
\end{equation}
We see that all 'interaction constants' (i.e. coefficients of the Taylor series expansion of $W$ in $\phi$) are uniquely determined from (\ref{50}).

Surely this oversimplified example has nothing to do with reality. However studying of more realistic options of application of self-consistency condition (\ref{48}) is beyond the scope of this article.

\subsection{IR brane stabilization and fixing mass hierarchy: an analogy with Coleman-Weinberg mechanism}

\qquad In the Randall-Sundrum model mass hierarchy is given by the ratio of locations of UV and IR boundaries (cf. (\ref{3})) of AdS space $\psi = \epsilon / L \approx 10^{-16}$. Regularization of IR divergencies of S-DW expansion of the induced one loop action with the introduction of the IR cut of AdS space at finite $z=L=M^{-1}_{SM}$ (see Sec. 4) makes one assume that stabilization of IR brane, i.e. violation of conformal symmetry $\psi \ne 0 $, may appear spontaneously in analogy with the Coleman-Weinberg mechanism of introduction of mass scale into a classically conformal theory thanks to the one loop quantum radiative corrections of potential \cite{Coleman}. We shall show that induced potential of Sec. 4 possesses corresponding non-trivial extremum.

Let us put down expressions for one loop induced potential in 4 dimensions as a function of $\psi$ in two cases: (1) when potential $V^{(4)}_{\it vac}$ is calculated with auxiliary mass method ($\alpha = \mu^{2}$ in (\ref{13})), and (2) when potential $V^{(4)}_{{\it vac}(GM)}$ is calculated as in Gubser and Mitra paper \cite{Mitra} ($\alpha = m^{2}$ in (\ref{13})):

\begin{equation}
\label{51}
V^{(4)}_{\it vac}(\psi) = \int_{\epsilon}^{L}\frac{dz}{(kz)^{5}}\,V^{(5)}_{\it vac} = \kappa \int_{\epsilon}^{L}\frac{dz}{z^{5}}\,\left(\frac{\sin \pi \nu}{16\pi^{3}}\right)\int_{0}^{\infty}y^{5}E(y, yL/z)\,dy,
\end{equation}

\begin{equation}
\label{52}
V^{(4)}_{{\it vac}(GM)} (\psi) = \int_{\epsilon}^{L}\frac{dz}{(kz)^{5}}V^{(5)}_{{\it vac}(GM)} = - \int_{\epsilon}^{L}\frac{dz}{z^{5}}\int_{0}^{\nu}{\tilde \nu}d{\tilde \nu}\left(\frac{\sin \pi {\tilde \nu}}{4\pi^{3}}\right)\int_{0}^{\infty}y^{3}E\,dy,
\end{equation}
\\
where $L= \epsilon / \psi$, $E$ is given in (\ref{43}), expression for $V^{(5)}_{\it vac}$ in (\ref{51}) is received if $F_{L}(z, \mu^{2})$ (\ref{42}) is used in the first term in S-DW expansion (\ref{25}) with subsequent integration over $\mu^{2}$ (cf. (\ref{29}) for $V^{(5)}_{\it vac}|_{L=\infty})$; and $V^{(5)}_{{\it vac}(GM)}$ in (\ref{52}) is given by expression (\ref{33}) for $V^{(5)}_{{\it vac}(GM)}|_{L=\infty}$ where substitution $K^{2}_{\nu}(y) \to E(y, yL/z)$ is performed.

Dependence of $E(y, yL/z)$ (\ref{43}) on $L$ is in $\gamma (yL/z)$ (\ref{41}) ($\omega L = yL/z$). Changing integration over $z$ in (\ref{51}), (\ref{52}) to integration over variable $v= yL/z$ and changing the order of integrations over $v$ and $y$ we receive finally following expressions for potentials (\ref{51}), (\ref{52}) as functions of mass hierarchy 'field' $\psi = \epsilon / L$:

\begin{equation}
\label{53}
V^{(4)}_{\it vac}(\psi) = \frac {V^{(5)}_{\it vac}(0)}{4 k^{5}\epsilon^{4}}\,(1- \psi^{4}) + \frac{\kappa}{64\pi\sin\pi\nu}\,\frac{\psi^{4}}{\epsilon^{4}}\,F_{1}(\psi),
\end{equation}

\begin{equation}
\label{54}
V^{(4)}_{{\it vac}(GM)}(\psi) = \frac {V^{(5)}_{{\it vac}(GM)}(0)}{4 k^{5}\epsilon^{4}}\,(1- \psi^{4}) - \frac{\psi^{4}}{\epsilon^{4}}\,\int_{0}^{\nu}\frac{1}{16\pi\sin\pi{\tilde \nu}}\,F_{-1}(\psi)\,{\tilde \nu}\,d{\tilde \nu},
\end{equation}
where $V^{(5)}_{\it vac}(0)$ and $V^{(5)}_{{\it vac}(GM)}(0)$ are given correspondingly in (\ref{29}) and (\ref{33}), and

\begin{equation}
\label{55}
F_{\lambda}(\psi)= \int_{0}^{\infty}\left[(\gamma(v)-1)\int_{\psi \cdot v}^{v} y^{\lambda}I^{2}_{\nu}(y)\,dy + \left(\frac{1}{\gamma(v)}-1\right)\int_{\psi \cdot v}^{v} y^{\lambda}I^{2}_{-\nu}(y)\,dy\right] v^{3}dv.
\\
\end{equation}

For $\psi \ll 1$ (observed value of $\psi$ is $\approx 10^{-16}$) $F_{\lambda}(\psi) = c_{1} + c_{2} \, \psi^{\lambda + 1 - 2\nu}$, thus potentials (\ref{53}), (\ref{54}) at $\psi \ll 1$ will have a form $V(\psi) = a + b [\psi^{4}-\alpha (\nu, r) \psi^{4+\delta}]$, where $\delta = 2(1-\nu)$ for $V^{(4)}_{\it vac}$ (\ref{53}) and $\delta = -2\nu$ for $V^{(4)}_{{\it vac}(GM)}$ (\ref{54}), and we showed here that coefficient $\alpha$ depends only on $\nu$ (i.e. on mass $m$ of scalar field) and on $r$ ('Robin coefficient' in boundary conditions (\ref{39})).

This form of potential resembles Coleman-Weinberg potential possessing non-trivial extremum at $\psi \ne 0$. Indeed derivative of potential $V(\psi) = a + b [\psi^{4}-\alpha (\nu, r) \psi^{4+\delta}]$ over $\psi$ is equal to zero at $\psi = 0$ and also at

\begin{equation}
\label{56}
\psi = \psi_{0} = \left[ \frac{4}{4 + \delta}\, \frac{1}{\alpha}\right]^{\frac{1}{\delta}}.
\end{equation}
Self-consistency of this expression demands $\psi_{0} \ll 1$. For small $\delta$ (i.e. for $\nu$ close to 1 for potential (\ref{53}) and $\nu$ close to 0 for potential (\ref{54})) the observed extra small value of mass hierarchy $\psi_{0}$ may be received even for moderate values of coefficient $\alpha (\nu, r)$. 

Corresponding calculations as well as the search for better justification of specific forms of induced potentials $V(\psi)$ in different models are the tasks for future. The main goal of this final remark is to demonstrate the existence of non-trivial extremum of potential calculated in the paper which may serve as a stabilization tool for IR-brane, fixing in this way the value of mass hierarchy.

\section*{Acknowledgements} Author is grateful for fruitful discussions and criticism to Andrei Barvinsky, Ruslan Metsaev, Dmitry Nesterov, Mikhail Vasiliev, Boris Voronov and other participants of Seminar in the Theoretical Physic Department of P.N. Lebedev Physical Institute.

\end{document}